\documentclass[12pt,epsf]{article}
\usepackage{graphicx}

\newcommand{\be}{\begin{equation}}
\newcommand{\ee}{\end{equation}}
\newcommand{\bea}{\begin{eqnarray}}
\newcommand{\eea}{\end{eqnarray}}
\newcommand{\beas}{\begin{eqnarray*}}
\newcommand{\eeas}{\end{eqnarray*}}
\newcommand{\ba}{\begin{array}}
\newcommand{\ea}{\end{array}}

\newcommand{\tr}{\mathrm{Tr}}

\newcommand{\nbox}{{\,\lower0.9pt\vbox{\hrule \hbox{\vrule height 0.2 cm \hskip 0.19 cm \vrule height 0.2 cm}\hrule}\,}}

\def\href#1#2{#2}

\begin{document}
\begin{titlepage}
\hfill

\vspace*{20mm}
\begin{center}
{\Large \bf Comments on quantum gravity and entanglement}

\vspace*{15mm}
\vspace*{1mm}

Mark Van Raamsdonk

\vspace*{1cm}

{Department of Physics and Astronomy,
University of British Columbia\\
6224 Agricultural Road,
Vancouver, B.C., V6T 1W9, Canada\\
{\it mav@phas.ubc.ca}
}

\vspace*{1cm}
\end{center}

\begin{abstract}
In this note, we attempt to provide some insights into the structure of non-perturbative descriptions of quantum gravity using known examples of gauge-theory / gravity duality. We argue that in familiar examples, a quantum description of spacetime can be associated with a manifold-like structure in which particular patches of spacetime are associated with states or density matrices in specific quantum systems. We argue that quantum entanglement between microscopic degrees of freedom plays an essential role in the emergence of a dual spacetime from the nonperturbative degrees of freedom. In particular, in at least some cases, classically connected spacetimes may be understood as particular quantum superpositions of disconnected spacetimes.

\end{abstract}

\end{titlepage}

\vskip 1cm

\section{Introduction}

Finding a general non-perturbative description of quantum gravity is surely among the most important and challenging open problems in all of theoretical physics. A central difficulty is that we do not even know the rules of the game. Are we allowed to assume that all of the basic tenets of quantum mechanics continue to hold when describing quantum gravity, or do we have to modify something fundamental, such as unitary evolution? Is spacetime a fundamental part of the description, or is it something that emerges only for special states? It is difficult even to know where to start when approaching these questions.

Significant progress has come in the past decade from string theory. The remarkable gauge theory / gravity correspondence \cite{bfss, malda,agmoo} posits that certain quantum field theories on fixed spacetime backgrounds are exactly equivalent to specific quantum gravitational theories in spacetimes with fixed asymptotic behavior. We can view this correspondence as providing a complete non-perturbative definition of the quantum gravity theory via a quantum field theory. At least for this set of examples, it appears that the physics of quantum gravity emerges from a completely conventional unitary quantum mechanical system (indeed, there are examples of the duality where the the field theory is nothing more than supersymmetric matrix quantum mechanics).

Given the progress, we can now ask a more refined set of questions: which conventional quantum theories have gravity duals? Which parameter values do we need to take, and which states should we choose so that a semi-classical spacetime description exists? How/why does spacetime/gravity emerge in these examples? Despite understanding many details about the correspondence, it is probably fair to say that we do not have a deep understanding how it works in general.\footnote{There is a general understanding that large N gauge theories in the 't Hooft limit give rise to string theories\cite{thooft}, and that quantized strings give rise to gravity. However, there are examples of gauge theory / gravity duality where the field theory is not a gauge theory in the 't Hooft limit and where the gravity side does not contain strings (e.g. M-theory on $AdS^4 \times S^7$).} Finally, how do we non-perturbatively describe quantum gravity for more general spacetimes? The known examples of this correspondence cover only a limited class of spacetimes (e.g. asymptotically AdS spacetimes or higher-dimensional asymptotically flat spacetime). For more general cases, such as cosmologically relevant spacetimes, it is not clear whether a holographic description should take the form of a conventional quantum mechanical system, or something more general.\footnote{Various proposals exist for the holographic description of de Sitter space or eternally inflating spacetimes (see e.g. \cite{strominger, gm, fssy,bf}), however these are at a somewhat preliminary stage of development. See also \cite{giddings} for a recent proposal of a general mathematical framework for quantum gravity.}

In this note, we will collect and discuss a number of known results and examples from string theory that we hope can provide some limited insight into these issues. The theme will be to focus on aspects of gauge-theory / gravity duality that relate to the most basic mathematical structure of the theories, namely the Hilbert space structure present in all quantum systems. We will argue that quantum-information theoretic ideas may be play an important role in understanding how quantum gravity arises in holographic descriptions.\footnote{Possible connections between quantum gravity and quantum information theory have been raised many times in the past. The full set of references would be too numerous to list, but we will refer to specific work in the main body of the paper.}

\begin{figure}
\centering
\includegraphics[width=\textwidth]{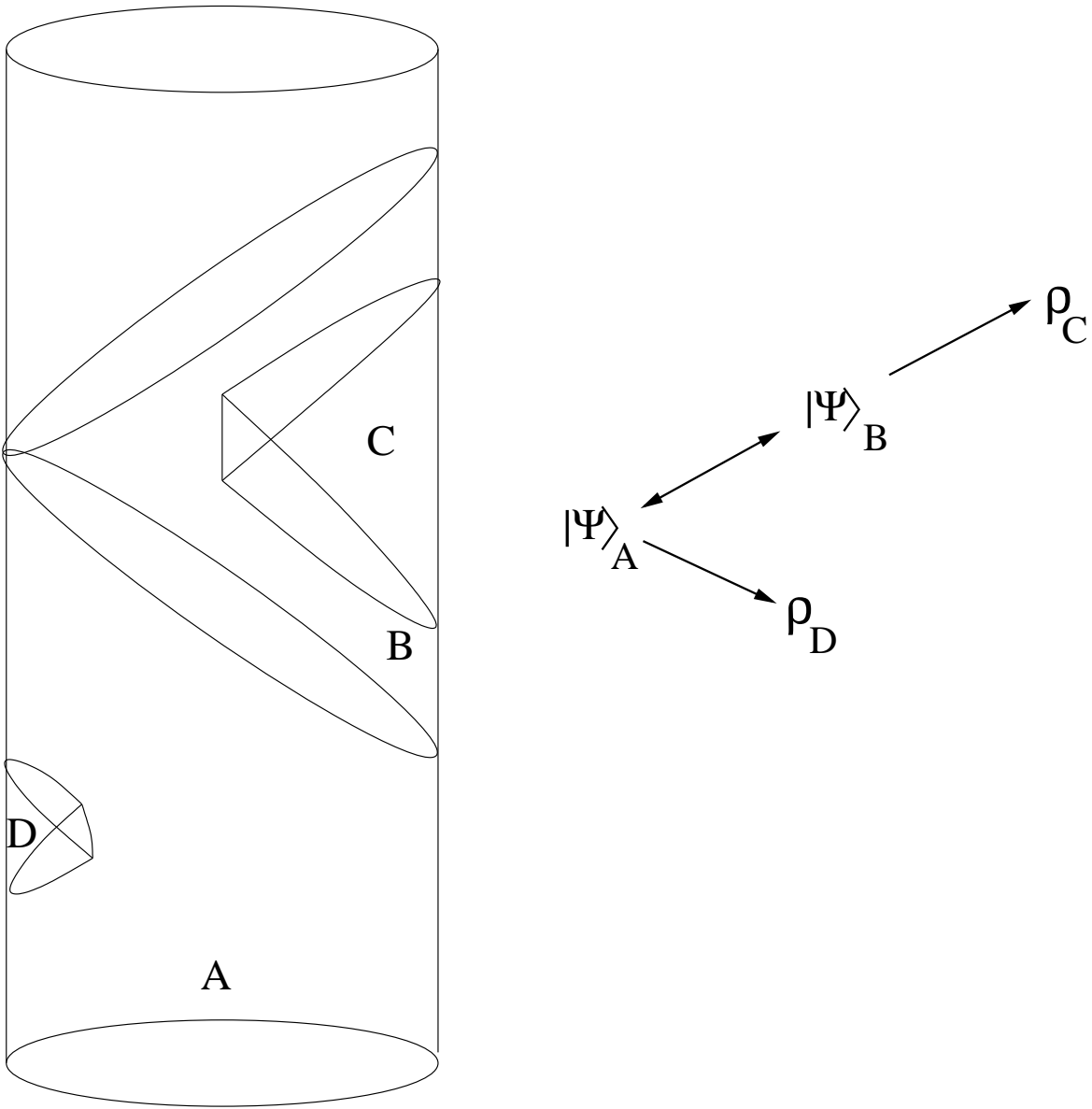}
\caption{Proposed mathematical structure of an asymptotically AdS quantum spacetime. For some choice of causal patches, we have a quantum systems associated to each patch and a set of maps between the spaces of states for the quantum system. For the example shown, a given spacetime can be associated with a state $| \Psi \rangle_A$ of the CFT on $S^d \times R$. This maps to a pure state of the CFT on $R^{d,1}$ associated with the Poincare patch $B$ and to mixed states of the CFTs on $H^d \times R$ associated with the smaller causal patches $C$ and $D$. See section 2 for details.}
\end{figure}

In section 2, we will offer a few observations relevant to the question of what a formulation of quantum gravity for general spacetimes might look like. We will argue that even in some familiar cases, the mathematical structure of quantum gravity is somewhat more general than in conventional quantum systems. Specifically, we note that in a number of examples of gauge theory / gravity duality, a complete quantum mechanical system or quantum field theory describes the physics in only a part of the corresponding spacetime, relevant to some particular set of coordinates or observers. The full description of a quantum spacetime may involve a number of different conventional quantum mechanical systems, just as a general manifold is described in terms of a number of different patches. Instead of transition functions mapping between different patches, we have maps between different Hilbert spaces. In some cases, these maps are isomorphisms, indicating that the descriptions are complementary to each other in the sense of black hole complementarity. More generally, a given quantum system contains only partial information about the entire spacetime. In this case, that system is generally in a mixed state, and the maps from other Hilbert spaces into the Hilbert space associated with these degrees of freedom are non-injective maps from states (or density matrices) to density matrices. While we do not attempt to extrapolate our observations beyond the known set of examples we discuss, it is possible that the type of structure we observe may be important in the holographic description of more general quantum spacetimes.

\begin{figure}
\centering
\includegraphics[width=0.5\textwidth]{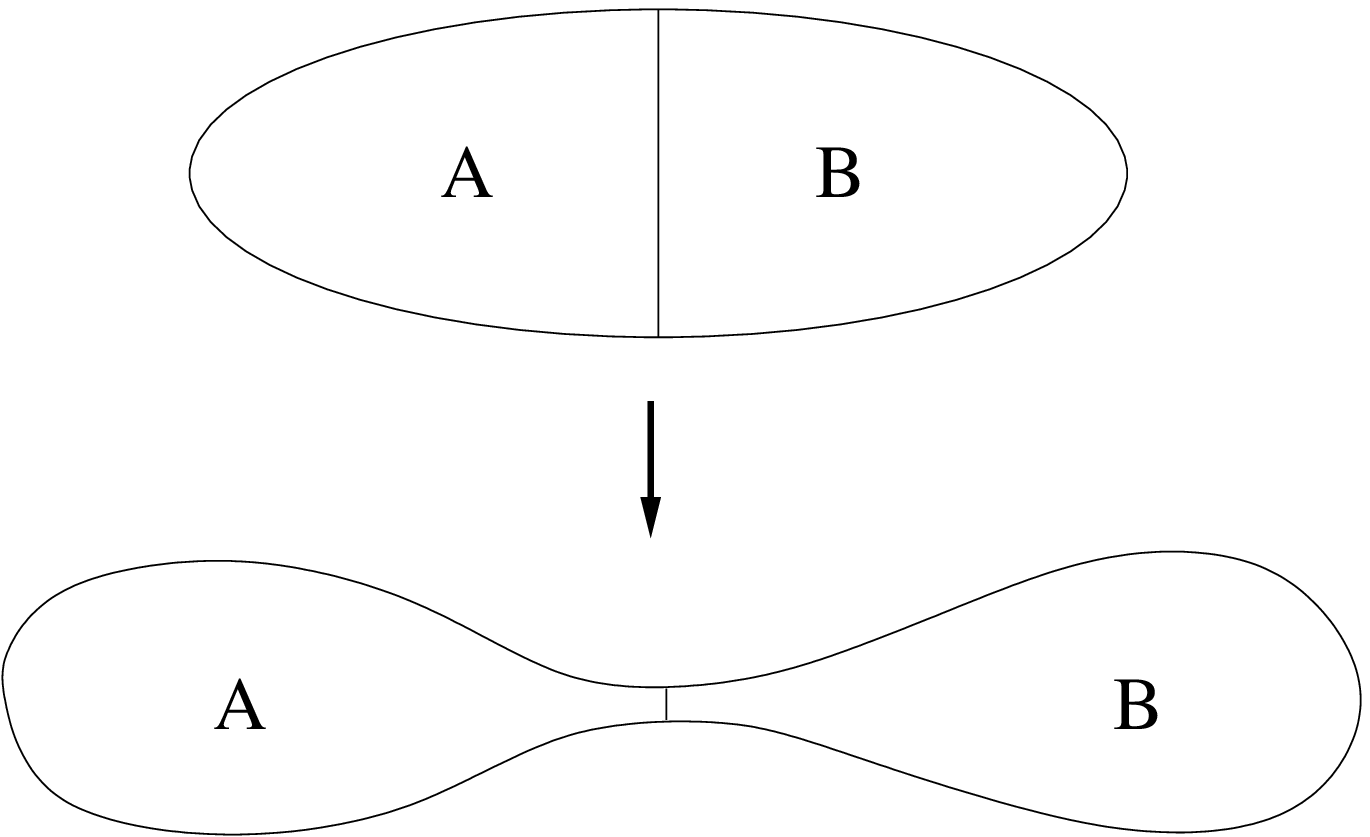}
\caption{Effect on geometry of decreasing entanglement between holographic degrees of freedom corresponding to $A$ and $B$: area separating corresponding spatial regions decreases while distance between points increases. }
\end{figure}

The discussion in section 2 suggests a patchwork structure for quantum spacetimes. In section 3, we will argue that the ``glue'' connecting various parts of spacetime together is quantum entanglement between the corresponding degrees of freedom in the non-perturbative description. We show that certain quantum superpositions of states corresponding to classically disconnected spacetimes may be interpreted as classically connected spacetimes. More quantitatively, we discuss several examples which suggest that decreasing the entanglement between two sets of degrees of freedom (e.g. by continuously varying the quantum state) effectively increases the proper distance between the corresponding spacetime regions, while decreasing the area separating the two regions. In other words, the two regions effectively pinch off from each other, as illustrated in figure 2.

Assuming this crucial role for entanglement in determining the structure of spacetime, we argue in section 4 that quantum Hamiltonians describing emergent gravity / spacetime should have low-energy states characterized by a large amount of entanglement. This property may be connected to the ``fast scrambling'' property that Sekino and Susskind \cite{ss} have argued is necessary for Hamiltonians describing black holes. While we would like to suggest that entanglement is necessary for emergent spacetime and gravity, we are certainly not suggesting that it is sufficient. Having the correct gravitational dynamics in the bulk (in particular, bulk locality) almost certainly requires Hamiltonians with other special properties beyond the ability to generate entanglement.

Finally, in section 5, we note that in some examples of gauge-theory / gravity duality (e.g. in cases where the field theory is just matrix quantum mechanics), it is a challenge to understand how to define precise measures of entanglement between the degrees of freedom, since there is not an obvious decomposition of the Hilbert space into a tensor product structure. Nevertheless, we argue that measures of entanglement should exist even for states in these theories. As an example, we show that even in a toy model 0+1 dimensional field theory, a physically motivated decomposition of the Hilbert space does exist, and the existence of this decomposition may be directly relevant to the emergence of a holographic radial direction.

We offer some concluding remarks in section 6.

\vskip 0.2 in
\noindent
{\bf Some related work}
\vskip 0.1 in

The mathematical structure that we observe in section 2 shares some features with an approach to quantum gravity called ``relational quantum cosmology'' \cite{rqc},\footnote{We thank Don Marolf for making us aware of this work.} which also involves associating quantum states in a number of different systems with a single quantum spacetime. The association of specific Hilbert spaces to particular causal patches is also implicit in Bousso's discussion of holography in general spacetimes \cite{bousso1, bousso2}, and it is central to the holographic space-time proposal of Banks and Fischler \cite{bf}. The present work is significantly less ambitious than these others in the sense that we are not attempting to present a complete mathematical framework for quantum gravity. Rather, we wish to point out some mathematical structure that already appears in concrete examples from string theory, in the hopes that this structure may be part of a complete framework that could apply to the description of more general quantum spacetimes. It would be interesting to understand better how well the specific structure we observe here fits into the various frameworks that have been discussed previously.

\section{What is quantum gravity?}

In this section, we will discuss a number of examples from string theory which suggest that the mathematical structure of quantum gravity may expand upon the structure present in ordinary quantum mechanics. In conventional quantum systems, states are vectors in some Hilbert space, and there is a canonical notion of time evolution (either of the state in the Schrodinger picture or of observables/operators in the Heisenberg picture) governed by a Hermitian Hamiltonian operator. These features are also present in the familiar examples of gauge-theory / gravity duality, but we will see (recall) that in this case, a particular Hilbert space with a particular Hamiltonian may give only a partial definition of the corresponding quantum gravity theory.

\subsection{CFT on $R^{d,1}$}

Let us begin by considering the most familiar formulation of AdS/CFT. It is believed that certain conformal field theories (e.g. ${\cal N}=4$ SYM theory) on $R^{d,1}$ are dual to gravity in asymptotically anti de Sitter space, with the CFT vacuum state dual to gravity in pure $AdS^{d+2}$ spacetime with metric
\be
\label{Poincare}
ds^2 = {R^2 \over z^2} (-dt^2 + dz^2 + d\vec{x}^2) \; .
\ee
Here, $z$ is restricted to $z>0$ and $z=0$ is the boundary of the spacetime. Evolution of the field theory state via the field theory Hamiltonian corresponds to evolution in the bulk according to the time $t$ (or more generally, a time which agrees asymptotically with $t$).

\begin{figure}
\centering
\includegraphics[width=0.25\textwidth]{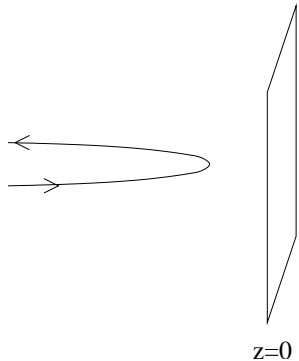}
\caption{Representative massive geodesic in Poincare-AdS spacetime.}
\end{figure}

\begin{figure}
\centering
\includegraphics[width=0.25\textwidth]{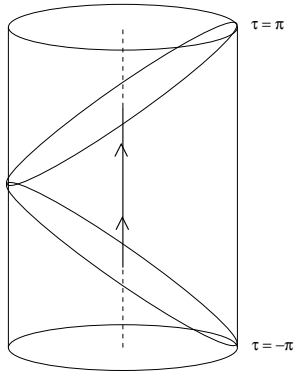}
\caption{The same geodesic in the global spacetime. The full evolution in Poincare time corresponds to evolution for a finite proper time, the solid part of the complete geodesic (which extends to $\tau = \pm \infty$).}
\end{figure}

Let us consider a representative massive particle geodesic in this spacetime, depicted in figure 3. The trajectory for one geodesic, related to the others by symmetry transformations, is
\be
\label{incomplete}
\vec{x} = 0 \qquad z(t) = \sqrt{R^2 + t^2}
\ee
with the particle reaching $z=\infty$ at $t = \pm \infty$. From the point of view of the field theory, the evolution of such a particle from $t=-\infty$ to $t=\infty$ in the spacetime corresponds to the complete evolution of a quantum state from negative infinite time to positive infinite time. However, from the spacetime point of view, it is easy to check that this evolution occurs in a finite amount ($\tau=R \pi$) of proper time for the particle. Of course, this corresponds to the fact that the spacetime defined by (\ref{Poincare}) is geodesically incomplete. More precisely, it defines only a part, the Poincare patch, of global AdS spacetime, with metric
\be
\label{global}
ds^2 = {R^2 \over \cos^2(\chi)} (- d \tau^2 + d \chi^2 + \sin^2 \chi d \Omega_{d}^2)
\ee
as shown in figure 4.

\begin{figure}
\centering
\includegraphics[width=0.25\textwidth]{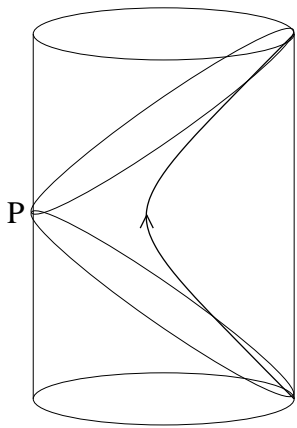}
\caption{Global AdS spacetime, showing the Poincare patch. This is the region inside the past and future light sheets emanating from a particular boundary point P, or the causal patch accessible to certain accelerated observers.}
\end{figure}

As shown in figure 5, the Poincare patch is the region inside the past and future light sheets from a particular boundary point, or alternatively, the causal patch\footnote{By causal patch, we mean the intersection of the causal past and causal future of the trajectory.} associated to an accelerated observer who begins and ends at the same boundary point at $\tau = \pm \pi$. In the global spacetime with coordinates (\ref{global}), the geodesic (\ref{incomplete}) corresponds to the part of the geodesic $\chi=0$ in the interval $\tau = (-{\pi \over 2}, {\pi \over 2})$ (see figure 4). From the global perspective, there is nothing special about this part of the geodesic; it is merely a part of a complete geodesic extending to $\tau = (-\infty, \infty)$.

To summarize, while the CFT on $R^{d,1}$ defines a complete conventional quantum system, the evolution of this system describes only a part of the evolution in the corresponding dual spacetime.\footnote{Another very explicit explicit example in which complete evolution in a quantum system corresponds to only a part of a dual spacetime has been given recently in the context of two-dimensional gravity by Karczmarek \cite{karczmarek}.} More precisely, since the boundary of the Poincare patch is a horizon for the accelerated observer, we can say that CFT on $R^{d,1}$ describes the physics outside the horizon for this observer. How do we study the physics ``behind the horizon''? In this case, the answer is well known: gravity in asymptotically AdS in global coordinates is dual to the CFT on $S^d \times R^1$. The evolution of a state in the field theory on $S^d$ should correspond to the complete evolution of the corresponding (geodesically complete) asymptotically AdS quantum spacetime.\footnote{Interestingly, the fact that the conformal field theory on Minkowski space is naturally related to a more complete theory defined on $S^d \times R$ was realized purely in a quantum field theory context in \cite{lm}, long before the advent of the AdS/CFT correspondence.}

So we have two different (though closely related) conventional quantum systems associated with the same spacetime.\footnote{For a more recent example of this, related to obtaining a description of the physics behind a black hole horizon, see \cite{hls}.} What is the precise relation between the two? Intuitively, knowing about the full spacetime should tell us everything about the physics in the Poincare patch. Thus, we expect that there is a (linear) mapping from states in the Hilbert space of the CFT on $S^d$ to states in the Hilbert space of the CFT on $R^d$. Of course, there is not just one Poincare patch in the spacetime, but a patch associated with any point on the boundary. So we really have a whole family of equivalent quantum systems, and for each one, we have a map from the Hilbert space associated with the CFT on $S^d$ to the Hilbert space associated to the specific Poincare patch.\footnote{This map should not be confused with the standard map between states of the CFT on $S^d \times R$ and local operators in the Euclidean CFT on $R^{d+1}$. Here we have a map between states and states. In the supergravity limit (e.g. {\cal N = 4} SYM in a limit where $N \to \infty$ and $\lambda \to \infty$), the Hilbert space of the CFT in either picture should be equivalent to the Fock space of supergravity fluctuations, so it should be straightforward to work out this mapping explicitly by transforming an arbitrary fluctuation mode in the global picture to Poincare coordinates and expanding in a basis of supergravity fluctuations more appropriate to the Poincare patch. Perhaps one could also understand explicitly what such a mapping looks like in the limit where the 't Hooft coupling goes to zero.}

For this example, it is probably true that these maps are isomorphisms. This is suggested by the fact that the global AdS and the Poincare patch share a Cauchy surface (e.g. the surface with $t=0$ or $\tau=0$ in our coordinates). So at least in the semiclassical limit, knowledge of the physics in a Poincare patch includes knowledge of the state of the various fields on a Cauchy surface for the entire spacetime, and this is enough to determine the entire future evolution in the global spacetime. So knowing the quantum state for the CFT on $R^{d,1}$ should allow us to reconstruct the state of the CFT on $S^d \times R$. Saying that the maps we have discussed are isomorphisms is in some sense a precise way saying that we have horizon complementarity \cite{stu} for the horizon associated with our accelerated observer. The quantum state of the CFT on $R^d$, i.e. the complete information about the physics outside the horizon, determines the quantum state of the CFT on $S^d$ and thus determines everything about the physics behind the horizon.

For the AdS space example, the quantum systems associated with this spacetime each describe the physics in the causal patch for a certain family of observers. It is natural to ask whether the physics accessible to an arbitrary observer defines some complete conventional quantum system, or at least a particular state in some Hilbert space. As an example, we recall (see figure 4) that a Poincare patch is the causal patch for some observer with a particular constant acceleration (parallel to velocity).\footnote{The choice of this observer is not unique, and there are also other observers with non-constant acceleration with the same causal patch.} Observers with a higher constant acceleration will have causal patches which are smaller in the sense that they are completely contained within a Poincare patch (but not the other way around). If there is a quantum system that describes the physics accessible to such an observer, we expect that knowledge of the state for the CFT on $S^3 \times R$ should give us all the information about what happens in the causal patch. Thus, there should again be a map from the Hilbert space of the CFT on $S^3 \times R$ to the Hilbert space of the quantum system dual to the causal patch. However, the states of this system will presumably not contain complete information about the full spacetime, since the causal patch for this observer does not include a Cauchy surface for the full spacetime. So in this case, the map cannot be an isomorphism.

\begin{figure}
\centering
\includegraphics[width=0.25\textwidth]{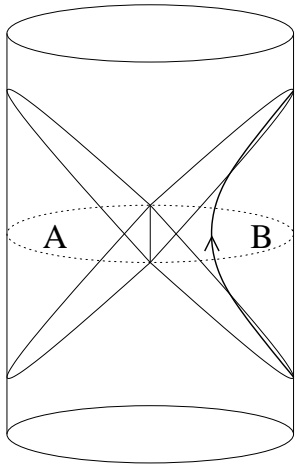}
\caption{Disjoint causal patches associated with a pair of accelerated observers. Neither patch contains a Cauchy surface for the whole spacetime, so the quantum systems associated with these observers will be in mixed states.}
\end{figure}

Furthermore, we do not expect such a map to generally take pure states to pure states, but rather pure states to density matrices.\footnote{See \cite{fhmmrs} for a discussion of mixed states in AdS/CFT, which inspired some of the ideas in this section.} This again is motivated by semiclassical considerations. As shown in figure 6, the causal patches A and B are analogous to two complementary Rindler wedges of Minkowski spacetime (see e.g. Birrell and Davies section 4.5). In that case, the Minkowski vacuum is perceived to be a thermal state by one of the Rindler observers; the perception of a mixed state corresponds to the fact that this observer does not have access to complete information about the spacetime. Correspondingly, it seems reasonable to expect that a pure state of the CFT on $S^d \times R$ should typically map to a mixed state of the quantum system associated with causal patches that do not include a full Cauchy surface. We will see another example of this case in the next subsection.

For the example of $AdS^3$, an interesting point is that the boundary of the causal patch for an observer with greater acceleration than the Poincare observer is again conformally equivalent to Minkowski space. Thus, the quantum system associated with such a patch should again be the CFT of $R^{1,1}$. The fact that the full spacetime maps to a mixed state of this CFT is perhaps the crucial feature that distinguishes such a system from one describing a Poincare patch of the same spacetime.\footnote{The idea that the physics of causal patches of $AdS^3$ smaller than the Poincare patch should be associated with mixed states of the Minkowski CFT has been previously recognized by Juan Maldacena \cite{juan}.}

For higher dimensional Anti de-Sitter spaces, the boundary of a causal patch smaller than the Poincare patch is conformally equivalent to $H^d \times R$ \cite{hls}, where $H^d$ is the topologically trivial space with constant negative curvature.\footnote{We thank Don Marolf for pointing this out.} So in this case, it is natural to conclude that the quantum system corresponding to the physics in such a patch is the CFT on $H^d \times R$. The magnitude of curvature of the $H^d$ does not matter here since we are talking about a conformal field theory, so the quantum systems associated with any causal patch smaller than the Poincare patch are all equivalent. What distinguishes one system from another in the complete description of some quantum spacetime is just the different (mixed) state that each field theory is in.

{\bf Summary:} In our example, we have seen that to a single quantum gravity theory, we can associate various different conventional quantum mechanical systems, most of which correspond only to a part of the spacetime (a specific causal patch). A particular quantum spacetime corresponds to a state (or density matrix) for each of the quantum systems. These systems have a partial ordering, such that for $A \ge B$, there is a linear map from the Hilbert space (or more generally, the set of density matrices) of A to the Hilbert space (or space of density matrices) of $B$ which determines the state of $B$ from the state of $A$. This is summarized in figure 1.

In this case, it happens that there is one description (the CFT on $S^d$) that is naturally associated to the complete global spacetime, but this may not be true in general. Of course, our discussion is quite analogous to the definition of a manifold in differential geometry. There, we define the whole manifold in terms of a set of coordinate patches, with maps between overlapping patches that indicate how the space is pieced together. In some cases, it is possible to describe the whole manifold with a single patch, but this is generally not possible.

\subsection{The eternal black hole in AdS}

In the example of the previous section, the CFT on $S^d \times R$ provides in some sense the most complete description of the dual gravitational physics, since evolution in the CFT maps to the full evolution in the global time coordinate. However, in the context of quantum gravity, even this system can, in other examples, form just a part of the description of a quantum spacetime, as we now recall.\footnote{See \cite{marolf} for another recent discussion which emphasizes that the CFT on $S^d \times R$ may describe a complete spacetime in some examples but only a part of a corresponding spacetime in other cases.}
\begin{figure}
\centering
\includegraphics[width=0.4\textwidth]{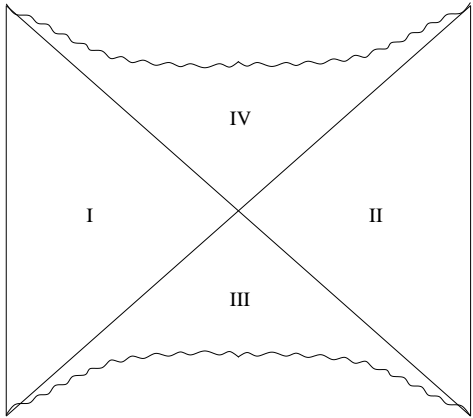}
\caption{Penrose diagram for the eternal AdS black hole spacetime.}
\end{figure}
Consider a Hilbert space which is the tensor product of two Hilbert spaces corresponding two copies of the CFT on $S^d \times R$. It has been argued \cite{eads,hm,bklt} that the pure state
\be
\label{hhstate}
|\psi \rangle = \sum_i e^{- \beta E_i \over 2} | E_i \rangle \otimes | E_i \rangle
\ee
in this Hilbert space corresponds to the eternal AdS black hole spacetime, whose Penrose diagram is sketched in figure 7. The motivation for this is as follows. First, the black hole spacetime contains two equivalent asymptotically AdS regions, suggesting that the dual description should involve two copies of the CFT. If we trace over the degrees of freedom associated with one of these CFTs, we obtain the density matrix
\[
\tr_2(|\psi \rangle \langle \psi |) = \sum_i e^{- \beta E_i } | E_i \rangle \langle E_i | = \rho_T \; ,
\]
the thermal density matrix for the other CFT. This is in accord with the expectation that the large AdS black hole spacetime should correspond to the thermal state in the corresponding conformal field theory.

Let us now see how this example fits into the framework of the previous section. In this case, we have a pure state (the ``Hartle-Hawking state'') in the tensor product Hilbert space that defines our quantum spacetime. We have a natural map between states in the tensor product Hilbert space and either of the two Hilbert spaces corresponding to our two CFTs. This map corresponds to tracing over the degrees of freedom of the other CFT, and generally maps a pure state of the tensor product Hilbert space to a density matrix for the reduced Hilbert space.
\begin{figure}
\centering
\includegraphics[width=0.4\textwidth]{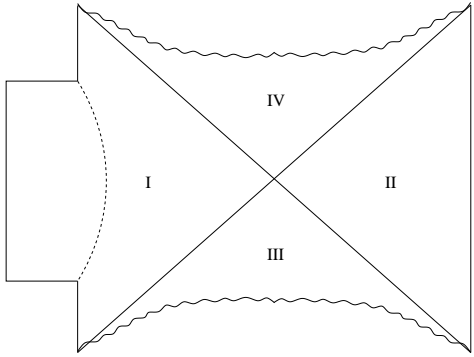}
\caption{Penrose diagram a spacetime identical to the eternal black hole in one asymptotic region but with a de Sitter bubble behind the horizon. Dashed line indicates a bubble wall (with tension chosen to satisfy junction conditions).}
\end{figure}
Note that this is a specific example where the maps we are discussing are not isomorphisms. Indeed, given a density matrix for one of the CFTs, we cannot reconstruct the whole spacetime. As discussed in \cite{fhmmrs}, it is easy to construct a spacetime which is identical to the eternal AdS black hole in one of the asymptotic regions outside the horizon but completely different behind the horizon. For example, as shown in figure 8, we can patch in a de Sitter bubble together with a positive tension bubble wall behind the horizon in place of the second asymptotic region. In this case, we expect that the asymptotic observer in the Anti-de Sitter region cannot distinguish between this new spacetime and the eternal black hole, so the density matrix should be the same.\footnote{Even in the previous example, there are many pure states in the tensor product Hilbert space that give rise to the same density matrix for one CFT. It seems reasonable to us that these correspond to states which differ in the second asymptotic region in such a way that they can't be distinguished by any experiments performed in the first asymptotic region. This contradicts a strong form of the complementarity principle which states that the region outside any horizon should contain complete information about the full spacetime. However, it does not conflict with a weaker statement that complementarity holds whenever the region outside the horizon contains a Cauchy surface for the entire spacetime.} In the de Sitter bubble case, it is not at all clear what the full description of the spacetime could be, since we do not know how to describe quantum gravity for asymptotically de Sitter space. However, it is possible that the whole spacetime corresponds to a state in {\it some} Hilbert space, and that the full definition of the quantum gravity theory includes a prescription for mapping states in this Hilbert space to the one associated with our CFT such that we obtain the thermal density matrix from the de Sitter bubble state.

In the preceding discussion, we have argued that there should be different spacetimes which correspond to precisely the same density matrix for a CFT describing one asymptotic region. An observer in this region cannot even in principle distinguish the two spacetimes. This is in contrast to another situation where we have {\it almost} indistinguishable spacetimes, namely when we are considering the microstates of a black hole. In this case, the microstates can in principle be distinguished from each other, or from the thermal state, as explained for example in \cite{bchlrs}.

\subsection{Combining theories with different Hamiltonians}

In the previous explicit examples, the quantum systems associated with the various parts of spacetime were all based on the same Lagrangian density. However, there can also be examples where the quantum systems are different.

For example, we can have spacetimes with two asymptotically AdS regions with different values for the cosmological constant. To see this, consider starting with the eternal AdS black hole spacetime. A spherically symmetric probe brane which starts out with infinite radius in one asymptotic region will have a collapsing trajectory as shown in figure 9. If the brane tension is sufficiently small, and if the brane sources a small change in the cosmological constant, there should be a corresponding back-reacted solution with essentially the same causal structure as in the diagram.
\begin{figure}
\centering
\includegraphics[width=0.4\textwidth]{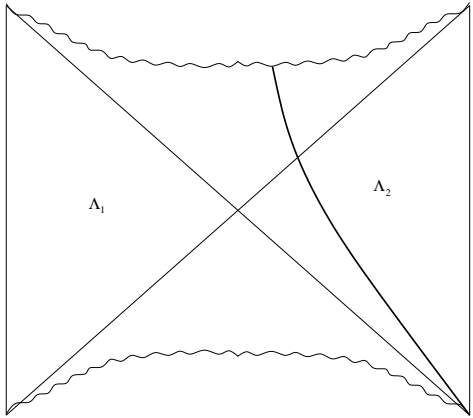}
\caption{Penrose diagram for a spacetime with two asymptotically AdS regions with different values for the cosmological constant. Bold curve is the trajectory of a collapsing spherical brane across which the cosmological constant changes value.}
\end{figure}
Since the cosmological constant differs in the two asymptotic regions, the holographic description should involve two different conformal field theories (e.g. with different gauge groups, or with different values for the central charge in 1+1 dimensional examples). We will discuss this case more below.

\section{Gluing with entanglement}

We have argued for a picture of quantum gravity in which quantum spacetimes are roughly analogous to manifolds, with the physics in particular patches of spacetime described by particular quantum systems. In this section, we will argue that quantum entanglement between the non-perturbative degrees of freedom corresponding to different parts of spacetime plays a critical role in connecting up the emergent spacetime.

\subsection{The eternal black hole again}

Consider first the example of a pair of CFTs on $S^3 \times R$. We reviewed above how a particular one-parameter family of pure states in the tensor product Hilbert space corresponding to the pair of CFTs is believed to correspond to the eternal AdS black hole spacetime (for various temperatures). It is natural to ask what other states in this Hilbert space correspond to. While we cannot give an answer in general, there is a clear interpretation for any product state
\[
|\Psi \rangle = |\Psi_1 \rangle \otimes |\Psi_2 \rangle \; .
\]
Since the degrees of freedom of the two CFTs do not interact in any way, and since there is no entanglement between the degrees of freedom for this state, the interpretation must be that we have two completely separate physical systems. If $|\Psi_1 \rangle$ is dual to one asymptotically AdS spacetime and $|\Psi_2 \rangle$ is dual to some other spacetime, the product state is dual to the disconnected pair of  spacetimes, as shown in figure 10.
\begin{figure}
\centering
\includegraphics[width=0.6\textwidth]{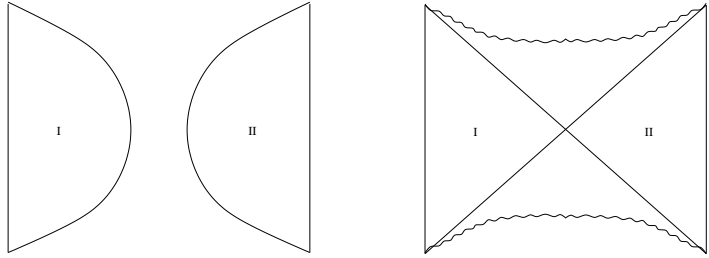}
\caption{Gravity duals for a product state and for an entangled state of two CFTs.}
\end{figure}
If product states correspond to completely disconnected spacetimes, it follows that any spacetime in which the two asymptotic regions are connected must be described by an entangled state.

Let us try to understand this connection in more detail. In an entangled state, a measurement on one part of the system affects the state of the other part of the system. In the eternal black hole spacetime (figure 7), the causal futures of the two asymptotic boundaries intersect behind the horizon. In particular, a probe sent into the horizon from asymptotic region 1 (which we can imagine as a localized disturbance arising from a measurement in the CFT${}_1$) enters the causal future of the second asymptotic region (region IV in the diagram). Complementarity suggests that at least some information about what goes in region IV is encoded in the state of CFT${}_2$, so the spacetime picture is consistent with the statement that the measurement of CFT${}_1$ affects the state of CFT${}_2$.

The same argument we have used here applies to the example in figure 9. In this case, we expect that the full spacetime corresponds to a state in a Hilbert space constructed as the tensor product of the Hilbert spaces for the two different CFTs. Since the two asymptotic regions are connected in the full spacetime (and, in particular, have intersecting causal futures), we expect that the state is one in which the degrees of freedom of the two CFTs are entangled. It would be interesting to understand more precisely how to construct such a state, but presumably it is similar to the Hartle-Hawking state (\ref{hhstate}) in the case where the back-reaction from the brane (and the change in cosmological constant) is small.

Our observations have the following interesting implication. Note that each term in the sum (\ref{hhstate}) is a product state, which we have argued corresponds to a spacetime with two completely disconnected parts. Thus, the ``literal'' interpretation of the Hartle-Hawking state is that it corresponds to a quantum superposition of many disconnected spacetimes. If it is true that the Hartle-Hawking state also corresponds to the eternal black hole spacetime, it must be that this connected spacetime is physically {\it the same} as an appropriate quantum superposition of disconnected spacetimes. In other words, it seems that the combined effect of a complicated quantum superposition may be captured by a simple classical spacetime. Of course, we do not expect simple superpositions such as
\[
|\Psi_1 \rangle \otimes |\Psi_1 \rangle + |\Psi_2 \rangle \otimes |\Psi_2 \rangle
\]
to have any effective description as a single classical spacetime, so it will be important to understand the degree and nature of the entanglement that is required for a superposition of product states to have an interpretation as a classically connected spacetime.

\subsection{Entanglement entropy}

We will now try to understand in a more quantitative way how entanglement of degrees of freedom relate to the connectedness of spacetime. Let us return to the case of a single CFT, defined either on the plane or the sphere. By locality, there are specific degrees of freedom in the field theory associated with any spatial region. Given some region A, the Hilbert space for the whole theory can be decomposed into a tensor product
\[
{\cal H} = {\cal H}_A \otimes {\cal H}_{\bar{A}} \; ,
\]
where $\bar{A}$ is the complement of $A$. For any state of the full quantum field theory, we can ask how the degrees of freedom in the region A are entangled with the degrees of freedom outside this region.

A simple quantitative measure of this entanglement is the entanglement entropy (see for example \cite{nc}). Given a pure state of the field theory, we can define the density matrix associated with any region $A$ as
\[
\rho_A = \tr_{\bar{A}}(|\Psi \rangle \langle \Psi |) \; .
\]
The entanglement entropy is defined to be the von Neumann entropy associated with this density matrix,
\[
S(A) = - \tr( \rho_A \log \rho_A)
\]
This is typically a divergent quantity, with the leading divergent piece proportional to the area of the boundary of $A$. However, we can consider a field theory defined with a cutoff (e.g. on a lattice), such that the entanglement entropy is finite.

\begin{figure}
\centering
\includegraphics[width=0.25\textwidth]{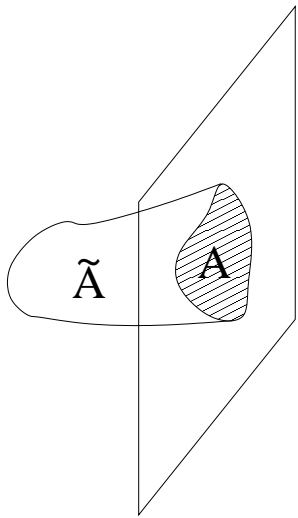}
\caption{Holographic interpretation of entanglement entropy.}
\end{figure}

Now, given some field theory with a gravity dual (e.g. strongly coupled large $N$ ${\cal N}=4$ SYM theory), we can ask what happens to the corresponding spacetime if we deform the state in such a way that the entanglement entropy associated with a region $A$ decreases toward zero (presumably at the cost of a large increase in energy). Thanks to Ryu and Takayanagi \cite{rt} (see also \cite{proof,hrt,hr,nrt}), we have an explicit proposal for the bulk interpretation of entanglement entropy: it is given by
\[
S(A) = {{\rm Area}(\tilde{A}) \over 4 G_N}
\]
where $G_N$ is Newton's constant, and $\tilde{A}$ is the minimal surface in the bulk such that the boundary of $\tilde{A}$ coincides with the boundary of $A$, as shown in figure 11. Very roughly, we can think of the surface $\tilde{A}$ as separating the region in the bulk described by the degrees of freedom in region $A$ with the region described by the degrees of freedom in the complement of $A$.

According to this proposal, if the entanglement entropy is decreasing to zero, the area of the minimal surface $\tilde{A}$ decreases to zero.\footnote{Strictly speaking, it is likely that the spacetime will cease to have a completely geometrical description before the entanglement entropy goes to zero, so our comments should be taken to hold in the regime where classical geometry is still a good description.} Since the surface $\tilde{A}$ is a dividing surface between two regions of the dual space, we see that as entanglement is decreased to zero, the two regions of space are pinching off from each other. Again, we conclude that entanglement is necessary for a bulk picture in which the two regions of spacetime are connected.

\subsection{Mutual information}

Another important quantum-information theoretic observable in a field theory is the notion of mutual information \cite{nc}. Generally, given any two sets of degrees of freedom in a quantum system, the mutual information is defined to be
\[
I(A,B) = S(A) + S(B) - S(A \cup B)
\]
This gives a direct measure of the entanglement between $A$ and $B$. It can be shown that $I(A,B)$ is always non-negative, and zero if and only if the density matrix for $A \cup B$ is the tensor product of the density matrices for $A$ and $B$.

Mutual information provides an upper bound on correlations in a system. It is straightforward to prove \cite{wvhc} that for any operators ${\cal O}_A$ and ${\cal O}_B$, acting only on the subsystems $A$ and $B$, we have
\be
\label{corrbound}
I(A,B) \ge \frac{(\langle {\cal O}_A {\cal O}_B \rangle - \langle {\cal O}_A \rangle \langle {\cal O}_B \rangle  )^2}{2|{\cal O}_A|^2 |{\cal O}_B|^2}
\ee
Thus, if we continuously vary a state such that the mutual information between $A$ and $B$ go to zero, then all correlations must decrease to zero also.
\begin{figure}
\centering
\includegraphics[width=0.25\textwidth]{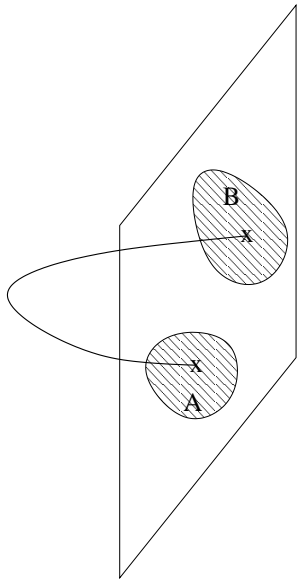}
\caption{Two point function of a local operator corresponding to a massive bulk particle. Length of connecting geodesic must go to infinity if mutual information between A and B decreases to zero.}
\end{figure}
Let us apply this to our CFT with a gravity dual. Consider any two disjoint spatial regions $A$ and $B$ of the field theory, and suppose we vary the state such that the mutual information between degrees of freedom associated with regions $A$ and $B$ goes to zero. Then for any local operator, the two point function of this operator with one point in region $A$ and the other point in region $B$ must go to zero. In particular, we can choose an operator corresponding to a very massive particle in the bulk. In this case, AdS/CFT tells us that the correlator is given by \cite{agmoo}
\be
\label{corr}
\langle {\cal O}_A(x_A) {\cal O}_B (x_B) \rangle \sim e^{-m L}
\ee
where $m$ is the mass and $L$ is the length of the shortest geodesic connecting $x_A$ and $x_B$ (again, we can work in a field theory with explicit cutoff so everything is finite). Combining (\ref{corr}) and (\ref{corrbound}), we see that as the entanglement between degrees of freedom in region $A$ and region $B$  (as measure by the mutual information) drops to zero, the length of the shortest bulk path between the points $x_A$ and $x_B$ must go to infinity.\footnote{Again, the geometrical picture of the dual spacetime probably breaks down before the mutual information goes to zero; our comments should be taken to apply only while the geometrical picture is still valid.}

Combining this result with the result of the previous subsection, we obtain the following picture. As the entanglement between two sets of degrees of freedom in a nonperturbative description of quantum gravity drops to zero, the proper distance between the corresponding spacetime regions goes to infinity, while the area of the minimal surface separating the regions decreases to zero. Roughly speaking, the two regions of spacetime pull apart and pinch off from each other, as shown in figure 2. Thus, there is a quantitative connection between entanglement measures and the structure of the dual spacetime.\footnote{In simple cases with some assumed symmetry, it has been shown that entanglement measures are enough to entirely reconstruct the dual spacetime \cite{hammersley,bilson}. It has also been suggested \cite{hubeny} that entanglement measures may be the optimal tool for such a reconstruction. Another proposal for how the entanglement structure of a state in a quantum field theory may relate directly to a dual spacetime has been given recently in \cite{swingle}.}

\subsection{The eternal black hole, take III}

In the context of a single CFT, it is difficult to come up with explicit examples where we can vary the quantum state, calculate the entanglement between degrees of freedom, and study what happens to the corresponding dual geometry. As usual, when the dual geometry is weakly curved, the field theory will be strongly coupled, and we have little hope of directly calculating entanglement entropy for any non-trivial states. On the other hand, we can do all of these things in our previous example of the eternal AdS black hole.

The entangled state of the doubled CFT
\be
|\psi \rangle = \sum_i e^{- \beta E_i \over 2} | E_i \rangle \otimes | E_i \rangle
\ee
was defined in terms of a parameter $\beta$ which gives the inverse temperature of the individual field theories. This parameter also controls the amount of entanglement between the two CFTs: as $\beta$ goes to infinity, the entanglement goes to zero, and we are left with a pure state
\[
|\psi \rangle = |vac \rangle \otimes |vac \rangle \; .
\]
More generally, the mutual information between the two CFTs is
\[
I(A,B) = 2S(TR)
\]
where $S(TR)$ is the entropy of the CFT on a sphere of radius $R$ at temperature $T$, proportional to $T^d$ for large $T$.

Let us now look at the dual geometry, taking the case of $AdS^3$ for concreteness. The metric in Kruskal-type coordinates is \cite{eads}
\[
ds^2 = \frac{-4}{(1+uv)^2}du \; dv + 4 \pi^2 T^2 \frac{(1-uv)^2}{(1+uv)^2}d \phi^2
\]
where $\phi \sim \phi + 2 \pi$ and we have set $R_{AdS}=1$. Defining coordinates $t$ and $x$ by $u=t+x$ and $v=t-x$, we can look at the geometry of the spatial slice at $t=0$ as a function of the temperature. The spatial metric is:
\[
ds^2 = {4 dx^2 \over (1-x^2)^2} + 4 \pi^2 T^2 {(1+x^2)^2 \over (1-x^2)^2} d \phi^2
\]
The circle at $x=0$ is the minimum area surface separating the two asymptotic regions, and we see that the area of this surface is
\[
4 \pi^2 T
\]
which decreases as the entanglement between the two CFTs decreases.

On the other hand, consider the length of the path between points in the two asymptotic regions where the $\phi$ circle has proper length $2 \pi r$. This length goes as
\[
L \sim 2 \ln \left( {\sqrt{{r \over 2 \pi T} + 1} + \sqrt{{r \over 2 \pi T} - 1} \over \sqrt{r \over 2 \pi T} + 1 - \sqrt{{r \over 2 \pi T} - 1}} \right) \; .
\]
We see that as entanglement between the two CFTs decreases to zero, this measure of distance between the two asymptotic regions goes to infinity, consistent with the general picture above. These results are summarized in figure 13, which provides an explicit example of the heuristic picture in figure 2.

\begin{figure}
\centering
\includegraphics[width=0.3\textwidth]{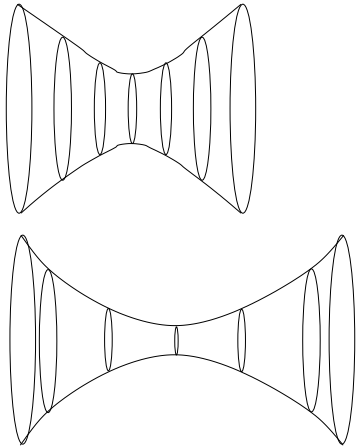}
\caption{Spatial section of eternal black hole for two different temperatures. For low temperature, where entanglement between the two CFTs is smaller, the asymptotic regions are further apart and separated by a surface of smaller area.}
\end{figure}

We should note that the present discussion probably should be restricted to the regime when the temperature is sufficiently large, $T > T_c \sim 1/R$, since below this critical temperature, the (Euclideanized) black hole geometry does not provide the minimum action configuration. For lower temperatures, the leading order $N^2$ part of the entanglement entropy associated with the Hartle-Hawking state drops to zero, since the field theory for these temperatures in in a confining phase.

\section{Generating entanglement}

The discussion in the previous section may give some insight into the question of which quantum systems have gravity duals. We have argued that quantum entanglement of the holographic degrees of freedom is essential for the appearance of a connected classical emergent spacetime. Of course, the degree of entanglement depends on the particular state we are considering, but in gauge theory / gravity duality, one most often considers the vacuum state (or other highly symmetric low-energy states). This suggests that Hamiltonians providing holographic descriptions of gravity in a dual spacetime should have low-energy states which are highly entangled.

In the context of two-dimensional conformal field theories, this assertion is easy to check. In the vacuum state of a theory with central charge $c$, the entanglement entropy between the degrees of freedom of an interval of length $L$ and the degrees of freedom outside this interval is 
\[
S(A) = {c \over 3} \ln \left( {L \over a} \right) \; ,
\]
where $a$ is a UV cutoff scale. Thus, the entanglement is directly proportional to the central charge of the CFT. On the other hand, in AdS/CFT examples, we also have \cite{agmoo}
\[
c \propto {R \over G_N}
\]
where $R$ is the AdS radius. The dual spacetime is classical (weakly curved in Planck units) only in conformal field theories with large enough central charge, or equivalently, those theories for which the vacuum state is sufficiently entangled. 

The notion that Hamiltonians for non-perturbative descriptions of quantum gravity should have highly entangled low-energy states may be connected to recent work of Sekino and Susskind \cite{ss} (following Hayden and Preskill \cite{hp}) who have argued that black holes must be ``fast scramblers''. The idea here is that to avoid a violation of the quantum no-xerox principle, black holes must be able to efficiently ``scramble'' quantum information. A pure state is scrambled if the density matrix for any subsystem is very close to a thermal density matrix unless the subsystem contains an order one fraction of the number of degrees of freedom of the system. If the scrambled state is perturbed (e.g. by placing a small subsystem in a pure state), the {\it scrambling time} is defined to be how long it takes for the system to return to a scrambled state. There is a fundamental lower bound on the scrambling time for a system based on the number of degrees of freedom, and Sekino and Susskind have argued that black holes (in flat spacetime) are systems that saturate this bound.

A thermal density matrix has the largest von Neumann entropy for all density matrices with a given energy expectation value,\footnote{To show this, we can simply extremize the action $S = -\tr(\rho \ln(\rho)) - \alpha(\tr(\rho) - 1)  - \beta(\tr(\rho H) - E)$ with Lagrange multipliers to ensure the correct trace and energy.} so we can think of a scrambled state as a state with energy equipartition where almost all subsystems are maximally entangled with the rest of the system. Thus, the ability of a Hamiltonian to quickly scramble a state is directly related to its ability to dynamically generate entanglement between its degrees of freedom. The statement that black holes are fast scramblers thus should translate to a statement that the Hamiltonian describing the fundamental degrees of freedom of a gravitational system should be very efficient at entangling these degrees of freedom.

It seems plausible that Hamiltonians which are fast scramblers should have low-energy states with lots of entanglement, but we do not know of any precise statement to this effect.

\section{Decomposing the Hilbert Space}

In the preceding discussion, we have focused on measurement of entanglement having to do with degrees of freedom of two separate field theories or two different spatial regions of a local quantum field theory. However, there are examples of quantum gravity theories for which the non-perturbative holographic description is a field theory defined at a point. For example, the BFSS matrix model and the Plane-Wave Matrix Model are ordinary quantum mechanical systems with Hamiltonians describing the interactions of a finite number of bosonic and fermionic matrices which are functions only of time. At first sight, it is difficult to see how to define any measures of entanglement here, since there is no obvious way to separate the Hilbert space into a tensor product structure.

Nevertheless, these systems contain many degrees of freedom, and we expect that it is still possible to define useful measures of how much these degree of freedom are entangled with each other. A useful hint may come from the fact that classically, the matrices in terms of which these models are defined can be related to noncommutative geometries. For certain matrix backgrounds in the large $N$ limit, these geometries become classical, and we end up with local field theories on these classical spacetimes. Since we can define sensible measures of entanglement for these limiting cases, it is reasonable to expect that such measures can be defined more generally.\footnote{It may be that a precise tensor product decomposition for the Hilbert space is not necessary to define measures of entanglement.}

We leave it as an open question to define useful measures of entanglement for matrix degrees of freedom that have a simple relation to the dual bulk geometry.\footnote{The recent work of Berenstein \cite{berenstein} on the emergence geometry from matrix degrees of freedom may be helpful here.} However, in the remainder of this section, we study as a toy model the simplest example of a 0+1 dimensional field theory and see that there is in fact a rather natural decomposition of the Hilbert space, and, correspondingly a set of observables related to the entanglement of degrees of freedom. We show further that this decomposition may relate directly to the structure on a dual emergent ``geometry'' (which is not particularly geometrical in this case).

\subsection{The harmonic oscillator / spin chain duality}

As our toy example, we consider the theory of a single massive scalar field in 0+1 dimensions, a.k.a. the harmonic oscillator. In this case, there is clearly no way to decompose the Hilbert space into a tensor product based on degrees of freedom at different spatial points. However, we note that there is a natural map between the states of a harmonic oscillator and the states of a semi-infinite spin chain. The energy eigenstates of the harmonic oscillator are denoted $|n \rangle$ where $n$ is a non-negative integer, and the state $|n \rangle$ has energy $E_n = \hbar \omega (n + {1 \over 2})$. A basis of states for the spin chain Hilbert space is given by
$$
|m_0 m_1 \cdots \rangle \equiv |m_0 \rangle \otimes |m_1 \rangle \otimes \cdots
$$
where we take $m_i = 1,0$ corresponding to whether the $i$th spin is up or down (i.e. $S^z_i$ eigenvalue $+\hbar/2$ or $-\hbar/2$). We can now define an isomorphism between the spin chain Hilbert space and the harmonic oscillator Hilbert space by identifying the basis elements as
\[
|m_0 \rangle \otimes |m_1 \rangle \otimes \cdots \leftrightarrow |n= m_0 + 2 m_1 + 2^2 m_2 + \dots \rangle
\]
That is, the state of the $j$th spin is determined by the $2^j$ digit in the binary representation of $n$.

The existence of such an isomorphism is not surprising: any two Hilbert spaces are isomorphic. However, we would like to argue that this particular mapping is natural and exhibits features consistent with the interpretation of the (discrete) direction along the spin chain as an emergent radial direction. As a first check, we note that the harmonic oscillator hamiltonian acts locally on the spin chain. It is straightforward to check that the Hamiltonian in spin chain language is
\begin{equation}
\label{spinham}
H = \omega \sum_{k=0}^\infty 2^k S^z_k + {\rm const}
\end{equation}
Physically, this is equivalent to a non-interacting semi-infinite chain of spins in a magnetic field that varies exponentially along the direction of the spin chain ($B(k) \sim 2^k$). Thus, the harmonic oscillator Hamiltonian does not induce any instantaneous long-range interactions, and the spatial ordering of the spins that we have chosen is natural, since the effective magnetic field $B(k) = 2^k$ increases monotonically with the site label $k$.

The isomorphism we have described amounts to a natural way to view the original harmonic oscillator Hilbert space as a tensor product of simpler Hilbert spaces. With this decomposition, we can define various measures of entanglement. By our criterion in section 6, the harmonic oscillator Hamiltonian is a particularly bad candidate to describe a classical emergent geometry, since its eigenstates are all product states in the spin chain picture. Nevertheless, there is a hint that this decomposition may be in some sense related to the emergence of a (non-geometrical) dual spacetime, as we will now see.

\subsection{The UV/IR correspondence}

In this section, we show that there is a natural relation between the spatial direction on the spin chain and energy scale in the field theory. We will call the $k=0$ end of the spin chain the infrared end of the radial direction and the infinite direction on the spin chain the ultraviolet.

Now, suppose we place an ultraviolet cutoff on the harmonic oscillator by restricting to states whose energy is $E - E_0 < \Lambda = \hbar \omega 2^n$.\footnote{This is slightly different than the usual type of ultraviolet cutoff used in field theory, which restricts momenta of individual modes, but an overall energy cutoff would also have the usual effect of restricting to low-momentum modes.} It is clear from the spin chain Hamiltonian (\ref{spinham}) that such a restriction corresponds precisely to restricting to the spins with $k<n$, since states with energy greater than or equal to $\hbar \omega 2^n$ are precisely those for which one of the spins with $k \ge n$ is excited (i.e. in the up state). Thus, an ultraviolet cutoff in the field theory corresponds to an infrared cutoff which truncates the spin chain from an infinite chain to a finite one. The radial position $k$ on the spin chain corresponds to an energy $\hbar \omega 2^k$ in the field theory.

In general, the number of spins in the dual spin chain is
\begin{equation}
\label{scales}
n \approx \ln \left( {\Lambda \over \omega} \right)
\end{equation}
We see that by decreasing the mass gap $\omega$, we also effectively increase the size of the spin lattice. This is also in accordance with standard AdS/CFT lore, where a mass gap in the field theory corresponds to a dual spacetime with an infrared end (as compared to the conformal field theory case where we have an infinite proper distance between any point in the spacetime and the infrared end).

To summarize, the relation between the harmonic oscillator and the spin chain includes a UV/IR correspondence completely analogous to the usual one from gauge-theory / gravity duality. This motivates an identification of the direction along the spin chain as a holographic radial direction associated with the harmonic oscillator. The moral of the story, which optimistically may carry over to quantum systems which are more relevant to quantum gravity, is that decomposing the Hilbert space in a way that may not have been initially obvious can provide direct insight into the emergence of a dual spacetime.\footnote{As an aside, we observe in appendix $A$ that considerations similar to the ones in this section provide some motivation for large $N$ gauge theories as natural candidates for theories with continuous dual spacetime as opposed to the discrete one that emerges here.} This example gives hope that there may be sensible measures of entanglement which are not related to the decomposition of the field theory degrees of freedom into groups associated with spatial regions.

\section{Discussion}

In this note, we have attempted to provide some insights into the structure of non-perturbative descriptions of quantum gravity using known examples of gauge-theory / gravity duality. We have argued that in familiar examples, a quantum description of spacetime can be associated with a manifold-like structure in which particular patches of spacetime are associated with states or density matrices in specific quantum systems. We have also argued that quantum entanglement between microscopic degrees of freedom plays an essential role in the emergence of a dual spacetime from the nonperturbative degrees of freedom. We hope that these insights might be useful in the efforts to formulate quantum gravity for more general spacetimes.

Perhaps the most fascinating aspect of this direct connection between quantum entanglement and geometry is that measures of entanglement are universal: they can be defined for any state in essentially any quantum system. In particular, these quantum-information theoretic observables give us a way to generalize the geometrical quantities to cases where no classical spacetime picture exists. Optimistically, it might be possible to define a ``gravity dual'' for an arbitrary quantum system by assigning a geometrical interpretation to measures of entanglement or other quantum-information theoretic observables. In cases where the system has a bone-fide conventional dual spacetime these observables should obey certain constraints (such as the ones discussed in section 3.5) and evolve in a way consistent with the evolution of the dual geometry by Einstein's equations (or more generally a gravitational effective theory including matter). Presumably, this happens only for extremely special theories (probably with large numbers of degrees of freedom, supersymmetry, strong coupling, etc...). However, there may be a larger class of theories where the dual picture could involve some generalized (e.g. discretized or noncommutative) version of geometry. If a generalized dual geometrical/gravitational interpretation could be attached to relatively broad class of quantum systems, one might have a more systematic way to understand why/how ordinary gravity emerges in the special cases where it does.\footnote{For a hint that entanglement measures may be given a geometrical interpretation even in very simple quantum systems, see appendix B.}


\section*{Acknowledgements}

We would like to thank Vijay Balasubramanian, Raphael Bousso, Veronika Hubeny, Hong Liu, Juan Maldacena, Don Marolf, Shiraz Minwalla, Mukund Rangamani, Moshe Rozali. for helpful discussions and/or comments on the draft. This work is supported in part by the Natural Sciences and Engineering Research Council of Canada, by and Alfred P. Sloan Foundation Fellowship, and by the Canada Research Chairs Programme.

\appendix

\section{Aside: the need for many degrees of freedom}

The discreteness of the emergent spatial direction in our toy model of section 5.1 is in contrast to the continuous spacetimes that we typically consider in gauge theory / gravity duality. However, it is well known (e.g. from lattice gauge theory) that continuous quantum systems can arise by taking suitable limits of discrete ones. Indeed, many approaches to quantum gravity involve some fundamental discreteness of space. To approximate a continuous spatial direction that is large compared to the Planck scale, we would require our discrete space to have a large number of sites. From (\ref{scales}), we see that the required ratio between the UV cutoff scale and the mass gap scale is exponential in the number of sites. Hence, obtaining a spatial direction of macroscopic length from our toy model (assuming we associate the spacing on the spin chain with the Planck scale), would require considering an unnaturally large range of energy scales in our field theory (e.g. a ratio of energy scales of $2^{10^{35}}$ if we want a meter of space).

Starting from our trivial example, it is interesting to ask how we might modify the theory to satisfy a ``naturalness'' criterion that a the separation of scales between the macroscopic scale and the Planck scale in the emergent spacetime to be similar to the range of energy scales in the field theory (rather than a logarithmic relation as we found for the single oscillator). For now, let us continue to consider ordinary quantum systems (0+1 dimensional quantum field theories) with a discrete spectrum of states. Again, we could label the states by $|n \rangle$ (ordered according to their energy, ignoring degeneracies) and use the same mapping to the semi-infinite spin chain Hilbert space.\footnote{ In general, this is much less natural than the harmonic oscillator case, since a generic field theory Hamiltonian will be very complicated in the spin chain picture. Furthermore, the most interesting examples of AdS/CFT based on 0+1 systems are dual to spacetimes with ten or eleven dimensions.  However, we are simply trying to get a crude picture of the type of energy spectrum that would lead to a macroscopic emergent radial direction with for a reasonable range of energy scales in the field theory.} In general, the relation between the number of spins (elements of the tensor product) and the range of energies in the field theory will then be
\[
\Lambda/E_0 = {E(2^{n}) \over E(0)} \; .
\]
If we want to avoid the logarithm that we found in (\ref{scales}), it is necessary that the density of states in the quantum mechanics grows exponentially, that is $\rho(E) \sim e^{c E}$ for some constant $E$, up to a prefactor. Remarkably, this type of growth is exactly what arises in the large $N$ limit of gauge theories\footnote{For example, if we have two matrices $A^\dagger$ and $B^\dagger$ of creation operators with energy $\omega$ transforming in the adjoint representation of a $U(N)$ gauge group, the number of gauge invariant states of the form $tr(A^\dagger B^\dagger B^\dagger \cdots A^\dagger)|0 \rangle$ with energy $n \omega$ is between $2^n/n$ and $2^n$, giving a density of states of roughly $exp(ln(2) E/\omega)$.}, which are precisely the theories that give rise to approximately classical gravitational dual theories.

While it is probably not wise to draw too many conclusions from a trivial toy model, it is intriguing that our simple observations suggest an explanation for why large $N$ gauge theories are natural candidates for theories with an emergent spacetime.\footnote{Of course, it is well known that string theories are naturally associated with large $N$ gauge theories in the 't Hooft limit \cite{thooft}, but our present considerations seem rather different.}

\section{A geometrical dual for qubits?}

In this appendix, we offer the following curious observation, perhaps a hint that measures of entanglement may have a geometrical interpretation more generally. Consider a system of three qubits. For any pure state of such a system, we can calculate the density matrix corresponding to any one of the spins, and each of these density matrices have an eigenvalue in the interval $[0,{1 \over 2}]$. The three eigenvalues $(\lambda_1, \lambda_2, \lambda_3)$ completely characterize the entanglement properties of the quantum state.\footnote{Note that the density matrix corresponding to a pair of spins has the same non-zero eigenvalues as the density matrix for the excluded spin.} We can now ask which triplets of eigenvalues are allowed. The answer \cite{hss} is that a quantum state with eigenvalues $(\lambda_1, \lambda_2, \lambda_3) \in [0,{1 \over 2}]^3$ exists if and only if the three eigenvalues form the sides of a triangle. Thus, there is a natural map between quantum states of a three-qubit system and triangles. While it would be silly to make any general conclusions based on this example, it is tantalizing that the complete information about the entanglement in this system can be naturally encoded in such a geometrical way.\footnote{For a system of $n$ qubits, the smaller eigenvalues $(\lambda_1, \dots, \lambda_n)$ of the density matrices for individual qubits obey a similar constraint: a quantum state giving a set of eigenvalues exists if and only if the eigenvalues form the $n$ sides of a polygon. For $n>3$, these eigenvalues don't completely characterize the entanglement in the system, since the eigenvalues of density matrices corresponding to larger subsystems give additional information.}

\end{document}